# First Person Singular

Stevan Harnad
Chaire de recherche du Canada en sciences cognitives
Institut des sciences cognitives
Université du Québec à Montréal
Montréal, Québec, Canada H3C 3P8

Department of Electronics and Computer Science
University of Southampton
Highfield, Southampton, United Kingdon SO17 1BJ

**ABSTRACT:** Brian Rotman argues that (one) "mind" and (one) "god" are only conceivable, literally, because of (alphabetic) literacy, which allowed us to designate each of these ghosts as an incorporeal, speaker-independent "I" (or, in the case of infinity, a notional agent that goes on counting forever). I argue that to have a mind is to have the capacity to feel. No one can be sure which organisms feel, hence have minds, but it seems likely that one-celled organisms and plants do not, whereas animals do. So minds originated before humans and before language -- hence, a fortiori, before writing, whether alphabetic or ideographic.

Among the questions of origin that have preoccupied our minds most, five stand out: the origin of the world, life, humankind, language and mind. Big-Bang theory is our current best bet on how the universe was born, about 14 billion years ago. Life on earth emerged from the primal soup when light polarized certain proteins and their structure became self-replicating, some 4 billion years ago. Our own species is more recent: about 300, 000 years old, based on the anatomy of our bodies, including our brains. Language itself leaves no traces. *Verba volunt, scripta manent* (and writing came too late in the day: about 10,000 years ago, a technological innovation rather than a bodily mutation). No one knows whether we could speak when we left our first fossils or artifacts. So language began somewhere between 300,000 and perhaps 50,000 years ago.

The origin of mind is the most vexed question of all. Some define our species as the talking ape, which would make language capacity part of our very essence, with us from our very first days. Is it conceivable that we could speak before we had minds?

According to some authors (such as my teacher, Julian Jaynes), our oral tradition (we might even call it our illiterate oral "literary" tradition, since it includes the songs and tales of Homer even before we had invented a way to write them down) might all have been mindless, the concept of "mind" having been invented or discovered quite late in the hominid day, much as the concept of "world," "life," "human," or "language" -- or, for that matter "origin" -- might all have come relatively late in the day.

But is having a mind the same thing as having a "concept" of mind? Am I a Zombie until I come up with the word "mind" to name what we refer to by that term? I am certainly not dead until I have a concept (let alone a word) for "life," and I surely have a world even before I name it, or inquire about when and how it began.

Not only is it unlikely that our species started out as mindless Zombies, but our predecessor species were not mindless either, any more than our contemporary cousins the apes are, even though they cannot speak. All of us, whether speakers or mute, including our pets, have mental states, just as surely as many of us see colors and all of us feel pain, even though some of us have no names or abstract descriptions for them – or for anything at all.

So if having a mind predates having a language, surely it predates having a written language. Yet according to Brian Rotman – formerly a mathematician, now a philosopher of technology – not only the mind, but other "ghosts" like God and the Infinite were born only after we had not only writing, but alphabetic writing. The road leading to this surprising conclusion is a rather complicated one, the critical factor, for Rotman, being embodiment – and disembodiment.

Back to origins: Not only is it still a matter of speculation *when* language began, but it is equally uncertain *how* and *why* it began: What were those dramatic Darwinian advantages that language conferred on our species, sufficient to shape our brains, relatively quickly, into what they are now, with their unique inborn ability and predisposition to acquire and use language, an ability every bit as biological as the bird's ability to fly, the fish's ability to swim, the eye's ability to see and the ear's ability to hear?

It is not that speculative hypotheses about language origins are lacking: It was the ease with which one could come up with the "bow-wow" theory, the "pooh-pooh" theory and the "yo-he-ho" theory that inspired the Société de Linguistique de Paris to ban the topic of language origins from the late $19^{th}$ to the late $20^{th}$ century. No, what is lacking is a Darwinian evolutionary scenario as compelling and credible as the ones we have for the origins of flying, swimming, seeing and hearing.

Here's one candidate: Maybe the way language helped us to survive and reproduce more successfully than other species was that it allowed us to transmit to one another by word of mouth what all other species have to learn the hard way, through individual, time-consuming, risky, trial and error experience. We will not settle here whether this was indeed the Darwinian advantage of language, or something else again. But whatever the advantage was, it had to have led, through evolutionary trial and error, to that radical genetic and physiological shaping of the language regions of our brain into what they are today, just as Darwinian advantages had shaped wings, fins, eyes and ears. So it seems quite natural to ask whether language originated directly in the form of spoken words, or it started out in some other bodily form.

Human beings who are born deaf today have the same language-specialized brains the rest of us have, but because they cannot hear, they use gestural languages -- of which there are many, just as there are many spoken languages. And like spoken languages, gestural languages are capable of "saying" anything and everything that can be said in any other language. It is important to understand, however, that gestural language is not pantomime. Some of its components may have originated in pantomime and practical acts but, exactly as in spoken language, the shape of its words is irrelevant insofar as their linguistic function is concerned, as Saussure stressed: The meanings of linguistic gestures do not reside in their resemblance to what they stand for, any more than those of spoken words do: "Mama" may well have originated from the movements and sounds of nursing, but that similarity is not relevant to its linguistic meaning and use; its shape might as well have been arbitrary, as most words are.

Brian Rotman, however, singles out and stresses the nonarbitrary, iconic shape of nonverbal gesture, as a means of depicting and expressing resemblance and emotion. He reminds us that this nonlinguistic expressive power of gesture is a consequence of its (likewise nonarbitrary) embodiment: It is the expressive power of bodily movement. It is also the depictive power of sensory images, which, as we all know, are worth much more than a thousand words. Rotman notes that with language, this sensorimotor and emotional expressive power is reduced, replaced instead by the symbolic descriptive power of words: "telling" instead of "showing." Spoken language still has tone of voice and other nonverbal accompaniments to supplement its expressive power, and gestural language retains even more of this nonverbal expressive potential. But, one can ask, is this nonlinguistic accompaniment still really necessary, now that we can tell all?

Written language proves that it is not. The mathematician, Alan Turing (to whom this topic owes more than a few of its fundamental insights) not only co-invented the computer but showed that it was universal, in that it could compute anything that was computable. Turing also designed the "Turing Test" whereby we try to ascertain whether a device has a mind by testing whether it can say and understand everything that a human being can say and understand. In other words, does the device have the full expressive (and understanding) power of language (including computation), as a human being does?

The Turing Test excludes the "body" of the candidate device, restricting all interactions to written ones, precisely because Turing did not consider those other, nonverbal expressive powers (*showing* rather than *telling*) to be essential to having a mind – or at least to testing whether a device has a mind. (He left it open whether the device might have to possess other capacities, nonverbal, embodied ones, not tested directly, but nevertheless needed in order to pass the verbal Turing Test. For example, if you wrote to the device "What does a sunset (or a smirk) look like?" it would not only have to draw upon the infinite number of words that a real person could use to describe what a sunset (or a smirk) looks like, but it would also have to be able to describe what it feels like to look at a sunset (or to see or produce a smirk). It is very possible that no device could do that – on a scale that was indistinguishable from a human being – if it had never seen a sunset and never felt what it feels like to see a sunset or to see and produce a smirk. These are embodied experiences and capacities.)

What the Turing Test exploits is the expressive power of disembodied language. This is the expressive power of arbitrary verbal symbols, divorced from the expressive power of nonverbal, bodily gesture. It is the power of symbolic *propositions* – with truth-values ("true" or "false") -- to say anything and everything. Showing, unlike telling, is neither true nor false. It is only if you "subtitle" it ("this is how he strangled her") that pantomime takes on truth value. But that is the truth value of the proposition (what is being told), not of the "this," which merely points to what is being shown. (Pointing has no truth value either; nor does emoting. So "expressive" really has two different meanings, one objective, formal and truth-valued, the other subjective, somatic and emotional: "feels meaningful to me.")

According to Rotman, language only became fully digitized, disembodied and divested of all residual analog properties when it became alphabet-based. (He calls this property "phonemic," which is curious, since phonemes are in fact the minimal meaningful acoustic/articulatory units of spoken language; he probably means "graphemic.") Only serially ordered, speaker-independent, written language from which even the residual iconicity and embodiment of ideographic writing systems like Chinese has been eradicated can give rise to certain "ghostly" (likewise disembodied) effects, such as the concept of a unitary mind, independent of the body, or the concept of a single, disembodied deity, or the abstract concept of infinity (consisting of the totality of things one can count, if one goes on counting forever). Rotman argues that (one) "mind" and (one) "god" are only conceivable, literally, because of (alphabetic) literacy, which allowed us to designate each of these ghosts as an incorporeal, speaker-independent "I" (or, in the case of infinity, a notional agent that goes on counting forever).

Rotman's arguments are largely hermeneutic, rather than analytic or empirical. We are invited to accept many interpretations, based largely on analogies and associations. (This use of written language seems, ironically, rather analog and impressionistic -- even verging sometimes on a private vocabulary: the reader will encounter many odd uses of words, such as "machinic," "monobeing," and "invisibilization"). Rotman seems to me to be right only about the formal concept of a completed infinity, which may indeed depend on first having a formal notational system, if "infinite" is to mean anything more than just the intuition that counting can go on and on.

The last part of the book is intended to be prophetic: Having transited from the preverbal world of sensorimotor gesture to the verbal and eventually alphabetic world whose disembodiment gave birth to the immaterial mind, godhead and infinity, we are today beginning, according to Rotman, to return, thanks to computer and network technology, to an increasingly "liquid" and virtual world that is more like somatic gesture than serial graphemes. The predicted effect will be that this virtual bodily reality will dissolve the alphabet-bred mind, which will move "beside itself" into a parallel, fragmented, distributed state rather like multiple personality disorder or the paradoxical state of "superposition" in quantum mechanics.

It is not obvious that this is a fate consummately to be wished for. The usual etiology of multiple personality disorder is early childhood trauma rather than spending too much time in front of a computer screen (although one now has students who, unlike the previous generations that had worried whether someone else might be a figment of their imaginations, now serenely contemplate the possibility that they themselves might be a figment of someone else's imagination – a part of their "virtual reality").

So yes, our minds and our senses and our sensory inputs can indeed play tricks on us. But Descartes probably put his finger on a firmer reality when he pointed out with his *cogito* (which is 1$^{st}$ person singular, not *cogitamus ergo sumus*!) that there are some things that one cannot doubt, as long as one is *compos mentis*: I can doubt that I have a body, but I cannot doubt that I have a mind, if by "mind" I mean (as I should) whatever it is that I (not "we") happen to be feeling at the moment. Things may not really *be* the way they feel, but they indubitably *feel* the way they feel – and feelings have only one feeler (even when the feeler is feeling plural). Virtual reality can alter *what* is being felt, but not *that* it is being felt. (It takes anaesthesia to do the latter, and that's not the kind of technology Rotman is talking about.)  Hence I "know" I am not a Zombie (nor multiple Zombies), and my prelinguistic predecessors knew it too, about themselves, even if they could not express it. So do our pets.

In a foreword to this book, Timothy Lenoir, Professor of History and Philosophy of Science and Technology at Stanford University, suggests that we can only perceive or know what someone else perceives or knows if we have an abstract symbolic representation of it, not only in language, but in writing. Current neural evidence suggests otherwise. Not only I, but illiterate, alalic moneys have "mirror neurons." These are active if and only if either I or you are in the same bodily state (e.g., gazing at a sunset, or smirking). We don't know how these neurons do it, but it's certainly not via language, let alone writing, and it's unlikely to be based on abstraction or reasoning, rather than a more elemental direct perception, as with most other things we perceive, such as size, shape, thrill and threat.

To have a mind is to have the capacity to feel. No one can be sure which organisms feel, hence have minds, but it seems likely that one-celled organisms and plants do not, whereas animals do. So minds originated before humans and before language -- hence, a fortiori, before writing, whether alphabetic or ideographic. Biological evolution altered bodies physically, shaping wings, fins, eyes, ears and eventually the brain basis of language capacity. Any further reshaping of our mental lives has so far been technological and informational rather than biological and somatic, including the invention and use of writing as well as computer technology. Technology may eventually reshape our bodies too, but that will be through physical, not virtual reality.